# Possible Electrochemical Origin of Ferroelectricity in HfO$_2$ Thin Films

Maya D. Glinchuk[1], Anna N. Morozovska[2\*], Yunseok Kim[3] and Sergei V. Kalinin[4 †]

[1] *Institute for Problems of Materials Science, National Academy of Sciences of Ukraine, Krjijanovskogo 3, 03142 Kyiv, Ukraine*

[2] *Institute of Physics, National Academy of Sciences of Ukraine,*

*46, pr. Nauky, 03028 Kyiv, Ukraine*

[3] *School of Advanced Materials Science and Engineering, Sungkyunkwan University (SKKU), Suwon 16419, Republic of Korea*

[4] *The Center for Nanophase Materials Sciences,*

*Oak Ridge National Laboratory, Oak Ridge, TN 37922*

**Abstract**

Recent observations of unusual ferroelectricity in thin films of HfO$_2$ and related materials have attracted broad interest to the materials and led to the emergence of a number of competing models for observed behaviors. Here we develop the electrochemical mechanism of observed ferroelectric-like behaviors, namely the collective phenomena of elastic and electric dipoles originated from oxygen vacancies formation in the vicinity of film surfaces, as well as from grain boundaries and other types of inhomogeneities inside the film. The ferroelectric phase is induced by the "electrochemical" coupling, that is the joint action of the omnipresent electrostriction and "chemical" pressure, which lead to the sign change of the positive coefficient $\alpha$ in the quadratic term $\alpha P^2$ in the order-disorder type thermodynamic functional depending on polarization $P$. Negative coefficient $\alpha$ becomes the driving force of the transition to the long-range ordered ferroelectric phase with the spontaneous polarization $P$ in the direction normal to the film surface. Using the above ideas, we estimated that the reversible ferroelectric polarization, as high as (0.05 – 0.2) C/m$^2$, can be induced by oxygen vacancies in HfO$_2$ films of thickness less than (20 – 30) nm. Semi-quantitative agreement with available experimental data is demonstrated.

---

[\*] Corresponding author: *anna.n.morozovska@gmail.com*
[†] Corresponding author: *sergei2@ornl.gov*

# I. INTRODUCTION

## A. Observation of ferroelectricity and ferromagnetism of HfO$_2$ thin films

Recent observations of ferroelectricity in HfO$_2$ thin films of and related solid solutions [1] has riveted the attention of scientific community both from fundamental perspective and due to potential for applications. For the latter, ferroelectricity in HfO$_2$ thin films enables application in ferroelectric memories due to ease of synthesis and compatibility with Si processing [2, 3, 4, 5]. From the fundamental perspective, this is the first example of the ferroelectric behavior in binary oxides, necessitating development of theoretical models of observed behaviors.

On the other hand the ferroelectricity in HfO$_2$ thin films has a crucial relevance for performance of ferroelectric memories [2, 6, 7]. It was found out that ferroelectricity in HfO$_2$ films can be associated with its Pca2 non-centrosymmetric orthorhombic phase [8], that appears due to influence of dopant atoms, film thickness, annealing conditions etc. [9, 10, 11]. In the majority of published papers (see e.g. [1-10, 12, 13, 14, 15]) thin films of HfO$_2$ and other binary oxides were doped with Si, Zr, Y, Al or Gd. Generally, observed ferroelectric behaviors manifested in relatively thin films and disappeared on transition for thicker samples. The ferroelectric response required wake-up by electric cycling, or thermal annealing in N$_2$ atmosphere, and generally ferroelectric behavior demonstrated complex time and bias history responses reminiscent of ferroelectric relaxors [16, 17].

A number of models for observed behaviors were proposed. Rabe et al [18] proposed a mechanism for intrinsic ferroelectric instability in ZrO$_2$ and related materials. Following this, Noheda group has recently shown that ferroelectric phase of HfO$_2$ has polarization in (111) orientation, this providing the possible explanation for complex polarization dynamics observed in (100) films [19]. However, in parallel to this intrinsic model for the ferroelectricity, a number of authors offer strong evidence on the role of vacancies on materials behavior, including the role of preparation and annealing conditions, possible reactions at electrodes, and general proximity of ferroelectric and electroresistive switching [20, 21, 22, 23]. For example, Polakowski and Muller [17] used conventional wake-up technology and did not observed ferroelectricity in the 20-nm thick undoped HfO$_2$ film, while the ferroelectric (**FE**) phase was observed for undoped HfO$_2$ film of thickness $h \sim 35$ nm by Nishimura et al. [16], who annealed the films in $N_2$ atmosphere. Note that Gd doped HfO$_2$ films prepared at different annealing conditions in $N_2$ atmosphere allowed to observe ferroelectricity for $h \leq 40$ nm [10], while without the annealing the thickness of the doped films with FE phase was about (10 – 20) nm.

Beyond ferroelectric and electroresistive studies, Venkatesan et al. [24] reported the unexpected magnetism in $HfO_2$ thin films on sapphire or silicon substrates, further giving rise to multiple observations and studies of so-called $d^0$-magnetizm [25, 26, 27, 28, 29, 30]. Extrapolated Curie temperature was much higher than 400 K, and the magnitude of magnetic moment and hysteresis loop characteristics were dependent essentially on the type of substrate. Investigations [26, 27] allowed providing insight on the nature of magnetic defects in oxides nonmagnetic in the bulk, i.e. on the mechanisms of $d^0$-magnetism. The quantum-mechanics calculations performed in Ref.[31] had shown that the ground state of neutral oxygen vacancy at the oxide interface is a magnetic triplet, so that the vacancies can induce magnetization in thin oxide films of $HfO_2$, $TiO_2$, $SnO_2$, etc. (see Ref. [32]). The disappearance of magnetization after annealing in oxygen atmosphere [30] confirmed the supposition that oxygen vacancies at the interface between the film and substrate are the main source of the magnetization [27]. Therefore the coexistence of ferromagnetic and ferroelectric phases indicates that the binary oxide thin films can be treated as multiferroics. Specifically, the room temperature structural ferroelastic phases in $HfO_2$ films include tetragonal, cubic or monoclinic symmetry phases [9, 16], creating the multi-component multiferroicity in the thin films.

Keeping in mind the important role of oxygen vacancies in the magnetization appearance in $HfO_2$, $TiO_2$ and $SnO_2$ films, here we seek to explore whether oxygen vacancies, which are known to be elastic dipoles, can be responsible for ferroelectricity origin in $HfO_2$ films, allowing for the fact that the surface-induced piezoelectric effect can transform elastic dipoles into electric ones. The theoretical calculations have been performed in Landau-Ginzburg-Devonshire (**LGD**) phenomenological theory approach. LGD theory applicability is valid in the mean field approximation, known as zero approximation of the self-consistent field theory. Keeping in mind the existence of FE phase with long range order at $T \leq T_C$ in thin films of binary oxides, it is obvious, that the mean field approximation is valid for a $HfO_2$ film at $T \leq 500$ °C. To take into account the correlation effects and polarization inhomogeneity in the film one has to use the first approximation of self-consistent field taking into account fluctuations of polarization. We will consider the order-disorder ferroelectric phase transitions, since oxygen vacancies (acting as electric dipoles) are randomly distributed in the film. On the other hand the oxygen vacancies are considered as elastic dipoles coupled with the polarization via electrostriction and Vegard effects (see **sections II.A** and **II.B**). The dependences of polarization on the film thickness $h$, oxygen vacancies concentration $N_S$ and temperature $T$ will be derived and analyzed in **section II.C**. The dynamics of polarization reversal and comparison of the calculated hysteresis loops of ferroelectric polarization and dielectric permittivity with experimental data [16, 17] will be presented in **sections III.A** and **III.B**, respectively. Discussion of obtained results and conclusion is presented in **sections IV** and **V**, respectively.

Keeping in mind that the impurities can lead to the increase of oxygen vacancies concentration because of excess charges compensation, the proposed mechanism can be applied to any other undoped or doped binary oxides.

## II. FERROELECTRICITY IN THIN HfO$_2$ FILMS
### A. The role of oxygen vacancies

Oxygen vacancies are typical defects in any oxide, and their concentration in a thin oxide film can be several orders of magnitude larger than in the bulk (see e.g. Ref.[33, 34]). In particular, the equilibrium concentration of oxygen vacancies in the bulk of homogeneous HfO$_{2-\delta}$ material is small enough, much less than $10^{26}$ m$^{-3}$, corresponding to the non stehiometry $\delta$ significantly smaller than 0.05. Notably that for high enough concentration of neutral/charged vacancies the possibility of the vacancy-ordered state appearance should be considered [35]; and the vacancy ordering can lead to the appearance of elastic/electric dipole sublattice induced by the interface.

The vacancies accumulation is strongly coupled with elastic field concentration [36]. In particular, the vacancies can further induce the FE phase in oxides via Vegard effect [37, 38, 39]. Here we denote the elastic dipole tensor of a vacancy as $W_{ij}$ [40] and their local concentration as $\delta N_V(\mathbf{r})$, so that additional elastic strain $u_{ij}^W(\mathbf{r})$ induced by vacancies is $u_{ij}^W(\mathbf{r}) = W_{ij}(\mathbf{r}) \delta N_V(\mathbf{r})$. The structure of elastic dipole tensor $W_{ij}$ is controlled by the local symmetry of the defect centre [40, 41, 42], and so it determines the elastic dipole orientation and anisotropic volume change in the vicinity of defect localization.

Theoretical calculations for HfO$_2$ performed in Ref.[10] had shown that the increase of pressure leads to the appearance of orthorhombic polar phases necessary for existence of FE phase. Since the Vegard pressure is proportional to the concentration of oxygen vacancies, the increase of the concentration can lead to the existence of FE phase in a broad range of thin films thickness. The role of the surface-induced piezo-chemical coupling should be discussed. Here, elastic dipoles originated from oxygen vacancies can be transformed into electric ones due to the surface-induced piezoelectric effect [43, 44] existing at distances less than (2-4) nm from the surfaces. The effect origin is the inversion symmetry breaking in the direction "z" normal to the film surface and corresponding electric dipole moment $p_3$ is proportional to the product $W_{i3}d^S_{3ij}$, where $d^S_{3ij}$ is the coefficient of surface-induced piezoelectric effect. However, the temperature dependence of the piezoelectric response is weak, since it is surface-induced and is not related with the ferroelectric soft mode [45]. Such electret-like polarization can shift hysteresis due to other mechanisms, but cannot switch it between several states.

Here we assume the presence of electric and elastic dipoles originated from oxygen vacancies in the film accumulated in the vicinity of surfaces and at the grain boundaries. The coupling between the dipoles stems from the Vegard effect, i.e. the elastic energy $\frac{s_{ijkl}}{2}\sigma_{ij}\sigma_{kl} + u_{ij}\sigma_{ij}$ contains the Vegard energy $\sigma_{ij}W_{ij}(\mathbf{r})\delta N_V(\mathbf{r})$ originated from the elastic dipoles induced by oxygen vacancies that couples with the elastic stress $\sigma_{ij}$ via electrostriction mechanism. Thus the main polarization-dependent contribution is $\sigma^P_{ij} = q_{ijkl}P_kP_l$, where **P** is the polarization induced by electric dipoles and $q_{ijkl}$ is the electrostriction strain tensor. For detailed calculations of FE phase it is important to discuss the transformation of the oxygen vacancies under transition from nonpolar phase to polar one. These allows us to consider the electric polarization **P** as a long-range order parameter, and include the conventional form of Vegard effect contribution in the thermodynamic potential [46, 47].

### B. Thermodynamic potential

The phonon modes in HfO$_2$ were found to be independent on temperature in wide temperature range (20 – 800) K [48], which means that static dielectric permittivity is also temperature independent and displacement-type soft mode models are not applicable for HfO$_2$ case. Treating oxygen vacancies as lattice defects, which are randomly distributed in the binary oxide films, we use a hybrid "order-disorder" and LGD type free energy functional for quantitative consideration of vacancy-induced ferroelectricity. Gibbs free energy density of oxide film with paraelectric nonlinearity has the following form [49, 50]:

$$G = \int_V d^3r \left( G_{PE} + G_{P\sigma} \right) + G_S, \tag{1}$$

where $G_{PE}$, $G_{P\sigma}$ and $G_S$ are the polarization, electrostriction and elastic, and surface contributions, respectively.

The free energy contribution $G_{PE}$ describing the ferroelectric order-disorder type transition can be expressed via the macroscopic polarization vector components, $P_i = P_d \left\langle \frac{p_{di}}{|p_{di}|} \right\rangle$, which are averaged over the orientation of elementary dipole moments $p_{di}$:

$$G_{PE} = -\frac{J}{2N_d}\frac{P_i^2}{P_d^2} - \frac{J_{nl}}{4N_d}\frac{P_i^4}{P_d^4} + \frac{k_B T}{2N_d}\left[\left(1+\frac{P_i}{P_d}\right)\ln\left(1+\frac{P_i}{P_d}\right) + \left(1-\frac{P_i}{P_d}\right)\ln\left(1-\frac{P_i}{P_d}\right)\right] \\ + \frac{\delta}{2}\frac{\partial P_i}{\partial x_j}\frac{\partial P_i}{\partial x_j} - P_i E_i \tag{2a}$$

The polarization $P_d \cong N_d p_d$, where $p_d$ is the concentration-independent value of elementary dipole moment. The constant value $N_d$ should be treated as an effective statistically averaged concentration of electric dipoles related (but not in a straightforward way) with the strongly inhomogeneous concentration $\delta N_V(\mathbf{r})$ of "frozen-in" or/and "dynamically" disordered oxygen vacancies sites. $J$ is effective interaction constant (the mean field acting on the each dipole from its neighbours) that can be zero or negative ($J \leq 0$) for the considered case of induced (improper) ferroelectricity. $J_{nl}$ is effective nonlinearity coefficient, the gradient coefficient δ>0 is determined by the correlation between dipole moments and by the contribution of the spatial gradients induced by the surface as well as by the inhomogeneities inherent to the film. For the interaction beyond nearest neighbour dipoles, the quantity δ can be several orders larger than the squared distance between dipoles.

Expression (2a) can be expanded in series on $P_i$ up to the sixth order [51]:

$$G_{PE} \approx \frac{\alpha}{2} P_i^2 + \frac{\beta}{4} P_i^4 + \frac{\gamma}{6} P_i^6 + \frac{\delta}{2}\frac{\partial P_i}{\partial x_j}\frac{\partial P_i}{\partial x_j} - P_i E_i. \tag{2b}$$

The LGD-expansion coefficients α, β and γ, renormalized by the pseudo-spin exchange, have the following form:

$$\alpha = \frac{N_d}{P_d^2}(k_B T - J), \quad \beta = \left(\frac{k_B T}{3} - J_{nl}\right)\frac{N_d}{P_d^4}, \quad \gamma = \frac{k_B T N_d}{5 P_d^6}. \tag{2c}$$

$E_i$ is the electric field that is the sum of applied field $E_i^{ext}$ and depolarization field $E_i^d$ originated from e.g. incomplete screening of polarization bound charge by imperfect electrodes (such as half-metallic TiN electrodes used in experiments [16, 17]). Boltzmann constant is $k_B = 1.38 \times 10^{-23}$ J/K.

The electrostrictive and elastic contributions to the free energy (1) are:

$$G_{P\sigma} = -Q_{klij}\sigma_{kl}P_i P_j - \frac{s_{ijkl}}{2}\sigma_{ij}\sigma_{kl} - u_{ij}^W \sigma_{ij}. \tag{3}$$

In Equation (3) $\sigma_{ij}$ is the elastic stress tensor, $Q_{ijkl}$ is the electrostriction stress tensor, $s_{ijkl}$ is the elastic compliances tensor. The summation is performed over all repeated indices.

The last term in Eq.(3) is the Vegard-type energy density

$$u_{ij}^W \sigma_{ij} = \sigma_{ij} W_{ij}(\mathbf{r})\delta N_V(\mathbf{r}), \tag{4a}$$

that is determined by electroneutral oxygen vacancies with concentration variation $\delta N_V(\mathbf{r}) \sim \left\langle \sum_k \delta(\mathbf{r} - \mathbf{r}_k) \right\rangle$. Since the main polarization-dependent part of the stress $\sigma_{ij}^P = q_{ijkl}P_k P_l$ is defined by the electrostriction contribution, Eq.(4a) yields:

$$u_{ij}^W \sigma_{ij} = q_{ijkl} W_{ij}(\mathbf{r}) P_k P_l \delta N_V(\mathbf{r}) + o(P_k^2, \delta N_V^2), \tag{4b}$$

Since the first term in Eq.(4b) is proportional to the second power of polarization, it renormalizes the coefficient $\alpha$ in the term $\frac{\alpha}{2}P_i^2$ in Eq.(2b) as $\alpha + q_{ijkk}W_{ij}(\mathbf{r})P_k^2\delta N_V(\mathbf{r})$. The function $o(P_k^2, \delta N_V^2)$ designates the small higher order terms.

Notably the defects, vacancies and ions, or their complexes, tend to accumulate in the vicinity of any inhomogeneities, surfaces, interfaces and grain boundaries [see e.g. **Fig.1(a)**], since the energy of vacancies formation at surfaces and around the inhomogeneities can be much smaller than in the homogeneous regions [52, 53, 54, 55].

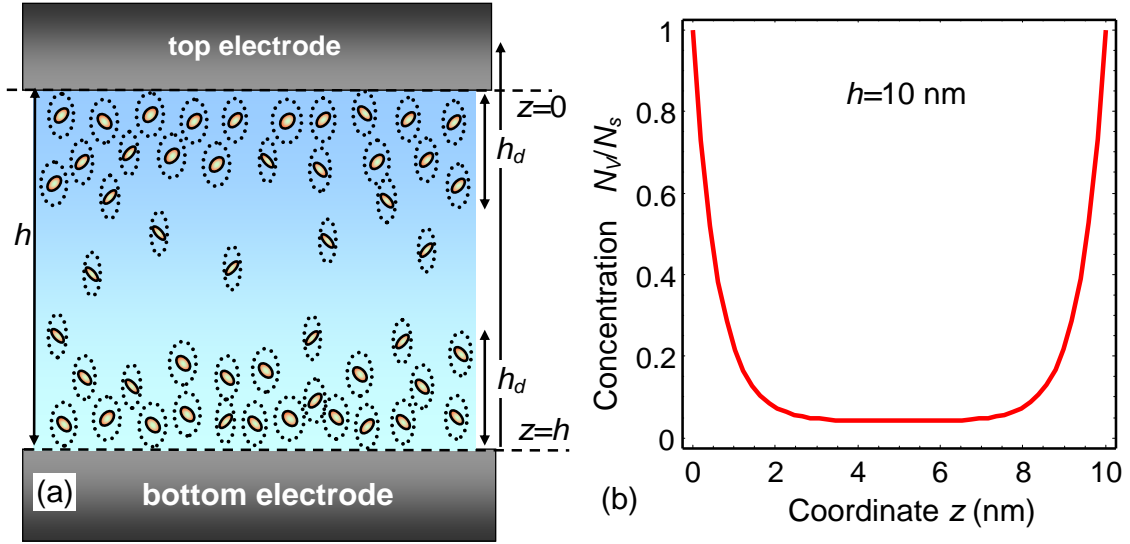

**FIGURE 1.** (a) Oxygen vacancies distribution in HfO$_2$ film of thickness $h$ placed between two equivalent electrodes. The vacancies concentration is maximal in the layers of thickness $h_d$ in the vicinity of the film-electrode interface. (b) The relative concentration of oxygen vacancies $N_V(\mathbf{r})$ corresponding to the distribution function (8a) in a 10-nm thick HfO$_2$ film with the vacancy-enriched layer of thickness $h_d = 0.6$ nm.

Defect accumulation can further lead to the emergence of strong electric and elastic fields, which in turn can lead to new phases appearance in oxides, for example, polar (ferroelectric) ones via percolation mechanism. On contrary, in the places where there are few vacancies the non-polar state remains. So, the polar ferroelectric and nonpolar states can coexist.

To describe these systems, we use classical LGD approach. Here, the equations of state $\partial G/\partial \sigma_{ij} = -u_{ij}$ determine the strains $u_{ij}$. Euler-Lagrange equations $\partial G/\partial P_i = 0$ determine the polarization components. Variation of the surface energy

$$G_S = \int_S d^2 r \left( \alpha_S P_3^2 + d_{3ij}^S P_3 \sigma_{ij} \right) \qquad (5)$$

along with the gradient terms in $G_{PE}$ and $G_{P\sigma}$ determines the boundary conditions,

$$\left. \alpha_S P_3 \pm \delta \frac{\partial P_3}{\partial z} + d_{3ij}^S \sigma_{ij} \right|_{z=0,h} = 0$$

($i$=1, 2, 3), which include so-called extrapolation length $\lambda = \dfrac{\delta}{\alpha_S}$, and contributions of the surface-induced piezoelectric effect proportional to $d_{3ij}^S$. The surface-induced piezoelectric effect exists in ultra-thin films [43, 44], even if the material is nonpiezoelectric (e.g. paraelectric) in a bulk.

It should be mentioned here that the surface energy did not include mismatch strains between the film and electrodes. The mismatch effect can lead to the appearance of the built-in electric field proportional to the product of surface-induced piezoelectric coefficient and mismatch strain (see [43, 44] and refs. therein). In the considered case of $HfO_2$ film between two symmetric electrodes the contributions of two build-in fields oriented in the opposite directions ("+z" and "–z", respectively) is assumed to compensate each other.

### C. Ferroelectricity induced by the oxygen vacancies

Putting elastic fields (4) into the Gibbs potential Eq.(1) leads to the renormalization of the coefficient $\alpha$ in (2c) $\alpha \to \alpha_{ij}^R$, namely:

$$\alpha_{kk}^R(\mathbf{r}) = \alpha_k^h + n_d + 2q_{ijkk}\sum_l W_{ij}\,\delta N_V(\mathbf{r} - \mathbf{r}_l). \qquad (6a)$$

The renormalization (6a) steams from the electrostriction coupling with Vegard expansion and depolarization field. The parameter $n_d$ is an effective depolarization factor originated from the depolarization field $E_{dep} \cong -n_d P_3$ directed normal to the film surface; and its value is determined by the electrode properties. Tensor $q_{ijkl} = Q_{ijmn} c_{mnkl}$ is the electrostriction strain tensor, $c_{mnkl}$ is the elastic stiffness. The intrinsic size effect [51] leads to the renormalization of coefficient

$$\alpha_k^h = \begin{cases} \dfrac{N_d}{P_d^2}(k_B T - J), & k = 1, 2, \\ \dfrac{N_d}{P_d^2}\left(k_B T - J\left(1 - \dfrac{h_0}{h}\right)\right), & k = 3. \end{cases} \qquad (6b)$$

where $h_0 = \dfrac{2\delta}{(\lambda + l_d)P_d^2}$ determines the correlation between the dipole moments and polarization extrapolation length $\lambda$. Parameter $l_d = \sqrt{J\delta/(4\pi N_d P_d^2 + k_B T - J)}$ is a characteristic length. Since $J \le 0$ the coefficient $\alpha_k^h > 0$ at all temperatures.

The statistical averaging of expression (6a) gives a nonzero value for the diagonal component:

$$\left\langle\!\left\langle \alpha_{kk}^R(\mathbf{r}) \right\rangle\!\right\rangle = \alpha_k^h + n_d + 2q_{ijkk}\left\langle\!\left\langle \sum_l W_{ij}\,\delta N_V(\mathbf{r} - \mathbf{r}_l) \right\rangle\!\right\rangle \cong \alpha_R + n_d + 2q_{ijkk}\left\langle W_{ij}\right\rangle \frac{1}{V}\int_V f_V(\mathbf{r})d\mathbf{r}. \quad (7)$$

Where $V = Sh$ is the film volume. The summation in Eq.(7) is performed over defect sites and the averaging of the function $\delta N_V(\mathbf{r} - \mathbf{r}_l)$ in the equation reduces to the integration over the layers, where the vacancies are accumulated. Hereinafter we assume that the distribution function $f_V(\mathbf{r})$ of vacancies $\delta N_V(\mathbf{r}-\mathbf{r}_l)$ depends on the distances from each of the film surfaces $z=0$ and $z=h$ (as the strongest inhomogeneity) in a symmetric way for a symmetric stack TiN/HfO$_2$/TiN, and reveals exponential decay far from the surfaces [see **Fig.1(b)**]. Namely:

$$f_V(\mathbf{r}) = \begin{cases} 0, & z<0 \\ N_S\left[\exp\left(-\dfrac{z}{h_d}\right)+\exp\left(-\dfrac{h-z}{h_d}\right)\right], & 0<z<h, \\ 0, & z>h. \end{cases} \quad (8a)$$

Here $N_S$ is the maximal value at the surface, $h_d$ is the decay factor of defect distribution function under the surfaces and $h_d \leq h$. Allowing for both surfaces, the average concentration of vacancies is

$$\overline{N}_V(h) = \frac{1}{V}\int_V f_V(\mathbf{r})d\mathbf{r} = 2N_S \frac{h_d}{h}\left(1-\exp\left(-\frac{h}{h_d}\right)\right). \quad (8b)$$

The value $\overline{N}_V(h)$ is proportional to $1/h$ for the realistic case $h_d \ll h$. The surface value $N_S$ is regarded independent on the film thickness, but depend on the film synthesis technology and annealing treatment. The treatment affect on the defect formation energy in accordance with e.g. Stephenson-Highland ionic adsorption model [56, 57]. Note, that the dependence of $N_S$ on the film thickness is not excluded for other models of sub-surface defect layer.

Putting Eqs.(7)-(8) in Eq.(6a) yields to the expressions:

$$\left\langle\!\left\langle \alpha_{kk}^R \right\rangle\!\right\rangle \cong \begin{cases} \dfrac{N_d}{P_d^2}(k_B T - J), & k=1,2, \\ \dfrac{N_d}{P_d^2}\left(k_B T - J\left[1-\dfrac{h_0}{h}\right]\right) + \dfrac{\Lambda/\varepsilon_0}{\varepsilon_{el}h+\Lambda\varepsilon_{film}} + 2q_{33}\overline{W}_{33}\overline{N}_V(h), & k=3. \end{cases} \quad (9)$$

Expressions (9) are valid for m3m symmetry. One can see from the expressions that the component $\left\langle\!\left\langle \alpha_{11}^R \right\rangle\!\right\rangle = \left\langle\!\left\langle \alpha_{22}^R \right\rangle\!\right\rangle \equiv \alpha$ remains positive, since $J \leq 0$. The positive depolarization factor $n_d = \dfrac{\Lambda/\varepsilon_0}{\varepsilon_{el}h+\Lambda\varepsilon_{film}}$ in expression for $\left\langle\!\left\langle \alpha_{33}^R \right\rangle\!\right\rangle$ originates from the depolarization field contribution due to the imperfect electrodes with finite screening length $\Lambda$ [58]. The local polar state emerging under condition $\left\langle\!\left\langle \alpha_{33}^R \right\rangle\!\right\rangle < 0$ is a priory not excluded in thin films for the case $q_{33}W_{33}<0$, when the vacancies concentration $\overline{N}_V(h)$ is high enough, exchange constant is small or zero, and the screening length $\Lambda$ of the electrodes are very small $\Lambda \ll 0.1$ nm (or zero in the limit of ideal electrodes).

The transition temperature to FE phase can be defined from Eq.(9) and has the form:

$$T_{cr}(h, N_S) = \frac{1}{k_B}\left[J\left(1 - \frac{h_0}{h}\right) - \frac{P_d^2}{N_d}\frac{\Lambda}{\varepsilon_0(\varepsilon_{el} h + \Lambda \varepsilon_{film})} - 2\frac{P_d^2}{N_d}q_{33}W_{33}\overline{N}_V(h)\right] \quad (10a)$$

Taking into account that $P_d \cong N_d p_d$, Eq.(10a) can be rewritten in the form:

$$T_{cr}(h, N_S) = \frac{1}{k_B}\left[J\left(1 - \frac{h_0}{h}\right) - N_d p_d^2 \frac{\Lambda}{\varepsilon_0(\varepsilon_{el} h + \Lambda \varepsilon_{film})} - 2p_d^2 q_{33}W_{33}N_d \overline{N}_V(h)\right] \quad (10b)$$

For the particular case $h_d \ll h$, zero pseudo-spin exchange constant $J = 0$ and ideally conducting electrodes with $\Lambda = 0$, the transition temperature $T_{cr}(h, N_S) = -\frac{2P_d^2}{N_d k_B}q_{33}W_{33}\overline{N}_V(h) \cong -\frac{4P_d^2}{N_d k_B}q_{33}W_{33}N_S \frac{h_d}{h}$ and so it is inversely proportional to the film thickness, $T_{cr}(h, N_S) \sim \frac{h_d}{h}$.

Note that Eqs.(10) contain nine fitting parameters, which are the screening length $\Lambda$, pseudo-spin exchange constant $J$, characteristic thickness $h_0$, polarization amplitude $P_d$, electric dipoles concentration $N_d$, oxygen vacancies distribution function with the amplitude $N_S$ and decay length $h_d$, and Vegard coefficient $W_{33}$. The parameters are unknown for HfO$_2$, but their typical ranges for other binary oxides, perovskite paraelectrics and order-disorder type ferroelectrics are well-established [33, 40, 42, 51]. Using the values of the fitting parameters for HfO$_2$, $J = 0$, $\Lambda = 0$, $P_d = 0.1$ C/m$^2$, $N_d \cong 5 \times 10^{24}$ m$^{-3}$, Vegard coefficient $W_{33} \cong -(10 - 20)$Å$^3$, $h_d = 0.6$ nm, and electrostriction constant $q_{33} = 5 \times 10^{10}$ m/F, we obtained from Eqs.(10) that $T_{cr}(h, N_S)$ can be higher that room temperatures (above 400 K) for the film thicknesses $h$ less than 15 nm. The estimate is in agreement with Nishimura et al [16] results, who revealed that the ferroelectric Curie temperature is about $T_C$=500 K for a 20 nm thick undoped HfO$_2$ film placed between two equivalent TiN electrodes.

Dependence of the transition temperature $T_{cr}(h, N_S)$ on the film thickness $h$ calculated for different surface concentrations $N_S$ are shown by curves 1 – 5 in **Fig.2(a)**. The transition temperature monotonically decreases with $N_S$ decrease (compare curves 1 – 5). For constant $N_S$ the transition temperature monotonically decreases as $1/h$ with $h$ increase. Namely $T_{cr} = 900$ K at $h=5$ nm, $T_{cr} = 300$ K at $h=14$ nm and $T_{cr} = 90$ K at $h=50$ nm for high $N_S = 10^{26}$ m$^{-3}$ [see curve 1 in **Fig.2(a)**]; meanwhile for smaller $N_S = 10^{24}$ m$^{-3}$ $T_{cr} = 10$ K at $h=5$ nm and $T_{cr} = 1$ K at $h=50$ nm [see curve 5 in **Fig.2(a)**]. Note, that the value $T_{cr} = 500°$C determined experimentally [10, 16] corresponds to HfO$_2$ film with thickness about 10 nm (see curve 2).

The value of the out-of-plane spontaneous polarization can be estimated as:

$$P_S(T,h,N_S) = \sqrt{\frac{\sqrt{\beta^2 - 4\gamma\langle\langle\alpha_{33}^R\rangle\rangle} - \beta}{2\gamma}} \approx \sqrt{-\frac{\langle\langle\alpha_{33}^R\rangle\rangle}{\beta}} \quad (11a)$$

For the case of the second order phase transition, parameters $\Lambda = 0$ and $J=0$ one gets from Eq.(11a) that

$$P_S(T,h,N_S) \sim \sqrt{T_{cr}(h,N_S) - T} \sim \sqrt{-\left(\frac{N_d p_d^2 \Lambda}{\varepsilon_0 \varepsilon_{el} k_B h_d} + \frac{2 p_d^2 q_{33} W_{33} N_d N_S}{k_B}\right)\frac{h_d}{h} - T}. \quad (11b)$$

The expression (11b) shows that the spontaneous polarization decreases as $\sqrt{\frac{B}{h} - T}$, with the film thickness increase.

Temperature dependences of the spontaneous polarization $P_S(T,h,N_S)$ calculated for the 10 nm HfO$_2$ film with different concentrations $N_S$ are shown in **Fig.2(b)**. At fixed temperature the spontaneous polarization monotonically decreases with $N_S$ decrease from $10^{26}$ m$^{-3}$ to $10^{24}$ m$^{-3}$ [compare curves 1 – 5 in **Fig.2(b)**]. For the definite $N_S$ value the polarization monotonically decreases with $T$ increase and disappears at $T>T_{cr}$ in accordance with the second order phase transition scenario. For $N_S = 10^{26}$ m$^{-3}$ the spontaneous polarization is as high as 0.2 C/m$^2$ at low temperatures, slightly higher that 0.1 C/m$^2$ at 300 K and disappears below 430 K [see curve 1 in **Fig.2(b)**]. The polarization is significantly smaller and disappears below 220 K for $N_S = 5\times10^{25}$ m$^{-3}$ [see curve 2 in **Fig.2(b)**]; is much smaller than 0.1 C/m$^2$ and exists only below 40 K for $N_S < 10^{25}$ m$^{-3}$ [see curves 3-5 in **Fig.2(b)**].

Temperature dependences of the spontaneous polarization $P_S(T,h,N_S)$ calculated for $N_S = 10^{26}$ m$^{-3}$ and different thickness $h$=(5 - 30) nm of HfO$_2$ film are shown in **Fig.2(c)**. The spontaneous polarization monotonically increases with $h$ increase [compare curves 1 – 5 in **Fig.2(c)**]. For the definite $h$ value the polarization monotonically decreases with $T$ increase and disappears at $T>T_{cr}$ in accordance with the second order phase transition scenario. For $h$=5 nm the spontaneous polarization is as high as 0.3 C/m$^2$ at low temperatures, slightly higher that 0.27 C/m$^2$ at 300 K and disappears below 830 K [see curve 1 in **Fig.2(c)**]; the polarization is higher than 0.15 C/m$^2$ at 300 K and disappears below 420 K for $h$=10 nm [see curve 2 in **Fig.2(c)**]; and exists only below (175 – 300) K for $30 \le h \le 15$ nm $N_S < 10^{25}$ m$^{-3}$ [see curves 3-5 in **Fig.2(c)**].

The crossing points of the curves 1-5 with temperature axes correspond to the transition temperatures at fixed concentration $N_S$ in **Fig. 2(b)** and fixed thickness in **Figs. 2(c)**, where the value for curve 2 for $h$ = 10 nm in **Figs. 2(c)** is close enough to the observed one [14]. One can see that the temperature dependence of polarization [shown in **Figs.2(b)-2(c)**] is $P_S(T) \sim \sqrt{T_{cr}(h, N_S) - T}$ in accordance with Eq.(11) for the second order phase transition scenario, where the transition temperature $T_{cr}$ increases with the film thickness decrease and concentration of oxygen vacancies increase.

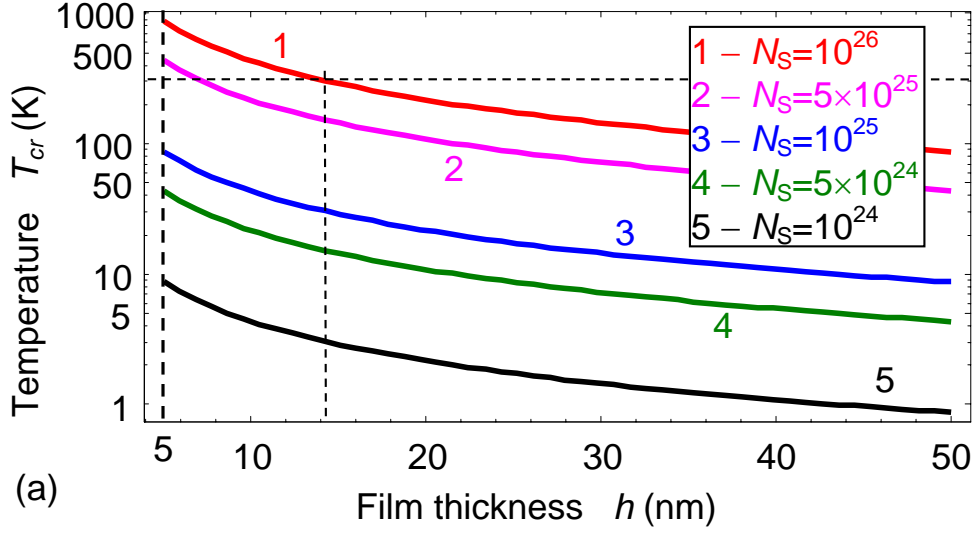

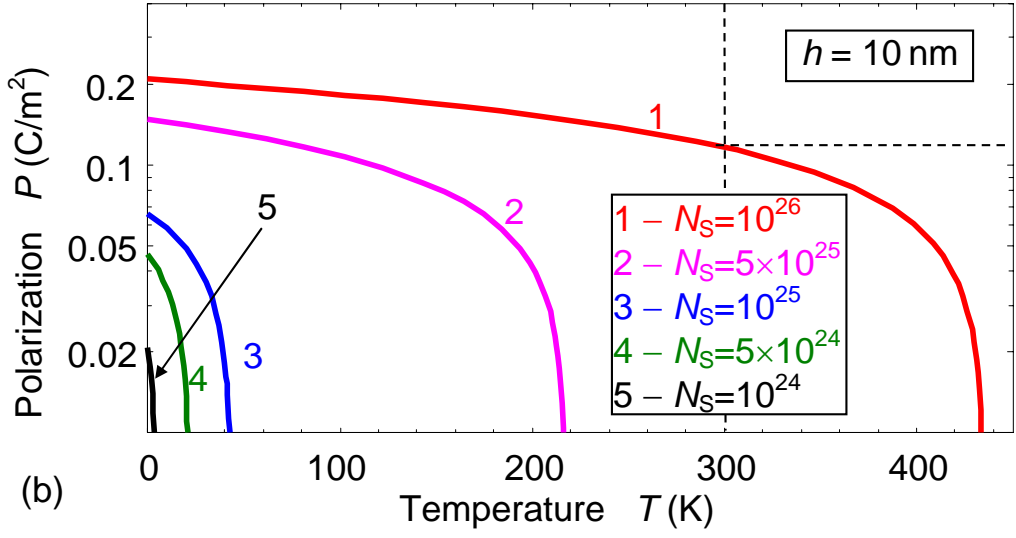

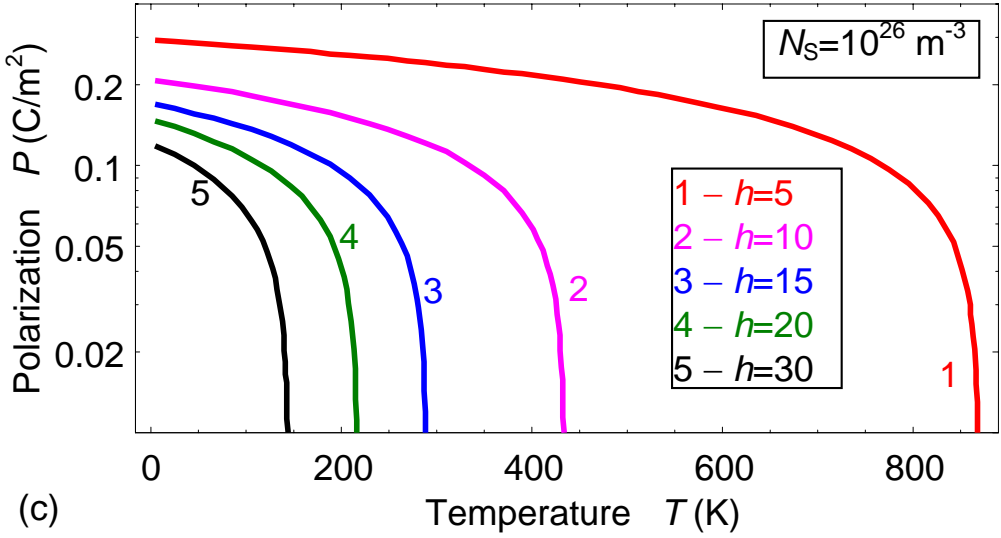

**FIGURE 2.** (a) Dependence of the transition temperature $T_{cr}(h, N_S)$ on the film thickness $h$ calculated for different maximal concentration of vacancies $N_S = (10^{26}, 5 \times 10^{25}, 10^{25}, 5 \times 10^{24}, 10^{24})$ m$^{-3}$ (curves 1, 2, 3, 4, 5). (b) Temperature dependence of the spontaneous polarization $P_S(T, h, N_S)$

calculated for $h$=10 nm and different $N_S = (10^{26}, 5 \times 10^{25}, 10^{25}, 5 \times 10^{24}, 10^{24})$ m$^{-3}$ (curves 1, 2, 3, 4, 5). **(c)** Temperature dependence of the spontaneous polarization $P_S(T, h, N_S)$ calculated for $N_S = 10^{26}$ m$^{-3}$ and different film thickness $h$=(5, 10, 15, 20, 30) nm (curves 1, 2, 3, 4, 5). Other parameters, $J = 0$, $J_{nl} = 0$, $h_0 = 0.4$ nm, $\Lambda = 0$, $P_d = 0.1$ C/m$^2$, $N_d \cong 5 \times 10^{24}$ m$^{-3}$, $q_{33} = 5 \times 10^{10}$ m/F, $W_{33} = -10$ Å$^3$, and $h_d = 0.6$ nm.

## III. DYNAMICS OF POLARIZATION REVERSAL

Nishimura et al [16] revealed that the ferroelectric Curie temperature is about Tc=500 K for a 20 nm thick undoped HfO$_2$ film placed between two equivalent TiN electrodes. The remanent polarization corresponding to the dynamic hysteresis loop measured at 100 kHz is about 0.25C/m$^2$ at room temperature. Dielectric response shows the double loop characteristic for a FE state. At the same time the coercive field ~(0.7 – 0.8) MV/cm appeared almost independent on the film thickness varying in the range (10 – 30) nm (see **Fig.1** and points in **Fig 3** in Ref.[16]).

Polakowski and Muller [17] observed that the remanent polarization strongly decreases from 0.1 C/m$^2$ for a 6-nm undoped HfO$_2$ film to 0.02 C/m$^2$ for a 12-nm film, and disappears for a 20 nm thick film at room temperature and frequency of applied field ~(1 – 30) kHz. Simultaneously the coercive field strongly decreases from ~1 MV/cm for a 6-nm undoped HfO$_2$ film to 0.2 C/m$^2$ for a 12-nm film, and disappears for a 20 nm thick film. Corresponding hysteresis loops of polarization (P-E) and dielectric response (C-E) were shifted along *E*-axis (see loops in fig.3 in Ref.[17]).

### A. Hysteresis dynamics of polarization and dielectric susceptibility

For a semi-quantitative comparison of the hysteresis loop observed in experiments [16-17] we model the out-of-plane polarization and susceptibility dependences on applied periodic electric field. The dynamics of the out-of-plane polarization dependence on applied periodic electric field $E(t) = E_0 \sin(\omega t)$ follows from the relaxation time-dependent equation

$$\Gamma \frac{dP}{dt} = \langle\langle \alpha_{33}^R \rangle\rangle P + \beta P^3 + \gamma P^5 - E(t). \quad (12a)$$

Hereinafter $\Gamma$ is the Khalatnikov coefficient. The "intrinsic" thermodynamic coercive field corresponding to the "bulk" analog of Eq.(12a) is equal to $E_C \cong \alpha_3^h \sqrt{\dfrac{27\alpha_3^h}{4\beta}}$, where the values of $\alpha_3^h$ are given by Eqs.(7) in the limit $h \to \infty$.

Dielectric susceptibility $\chi(T,h,N_S)$ can be calculated as a polarization derivative on electric field, $\chi(T,h,N_S) = \dfrac{dP(T,h,N_S,E)}{dE}$, and obeys the time-dependent equation

$$\Gamma \frac{d\chi}{dt} = \langle\langle \alpha_{33}^R \rangle\rangle \chi + 3\beta P^2 \chi + 5\gamma P^4 \chi - 1. \qquad (12b)$$

Typical dependences of polarization $P_S(T,h,N_S)$ and dielectric susceptibility $\chi(T,h,N_S)$ on applied voltage calculated for $h$=10 nm, room temperature and different maximal concentration of vacancies at the surface $N_S = (10^{26} - 10^{24})$ m$^{-3}$ are shown in **Fig.3(a)** and **Fig.3(b)**, respectively. The loop opening rapidly decreases with $N_S$ decrease from $10^{26}$ m$^{-3}$ to $10^{24}$ m$^{-3}$ and the remanent polarization changes from 0.75 C/m$^2$ for $N_S = 10^{26}$ m$^{-3}$ to 0.12 C/m$^2$ for $N_S = 10^{24}$ m$^{-3}$ [compare loops 1 – 5 in **Fig.3(a)**]. The polarization loop shape changes from the squire-like one with high enough coercive field for $N_S = 10^{26}$ m$^{-3}$ to a rather slim and tilted loop for $N_S$ decrease up to $10^{24}$ m$^{-3}$ [compare loops 1 – 5 in **Fig.3(a)**]. The dielectric susceptibility loop shape changes from the butterfly-like shape with two pronounced high maxima, which are well-resolved for $N_S = 10^{26}$ m$^{-3}$, to the slim loop with rather low and smeared maxima, which are hardly resolved with $N_S$ decrease to $10^{24}$ m$^{-3}$ [compare loops 1 – 5 in **Fig.3(b)**]. Further decrease of $N_S$ does not lead to any changes in the loop shape and sizes, but for $N_S$ smaller than $10^{25}$ m$^{-3}$ the loop itself is the dynamic effect, and the frequency decrease leads to the disappearance of the loop.

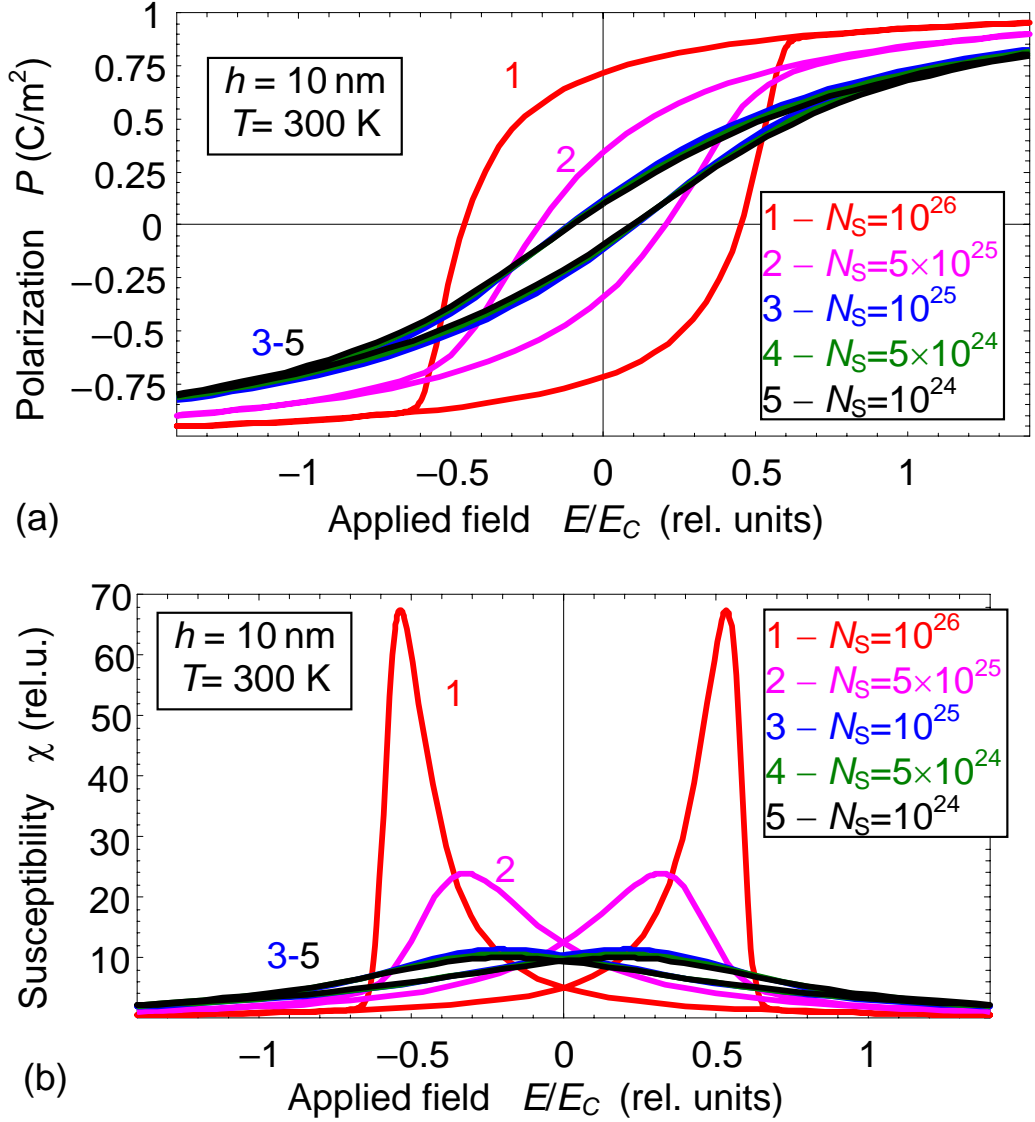

**FIGURE 3.** Polarization $P_S(T,h,N_S)$ (a) and dielectric susceptibility $\chi(T,h,N_S)$ (b) dependences on applied voltage calculated for $h$=10 nm, room temperature $T$=300 K and different concentration of vacancies at the surface $N_S = (10^{26}, 5\times10^{25}, 10^{25}, 5\times10^{24}, 10^{24})$ m$^{-3}$ (curves 1, 2, 3, 4, 5). Dimensionless frequency $w$=0.05. Other parameters are the same as in **Fig.2**.

Polarization $P_S(T,h,N_S)$ and dielectric susceptibility $\chi(T,h,N_S)$ dependences on applied voltage calculated for room temperature, fixed $N_S = 5\times10^{25}$ m$^{-3}$ and different film thickness $h$=(10 - 30) nm are shown in **Fig.4(a)** and **Fig.4(b)**, respectively. The loop width decreases in more than 2 times with $h$ increase from 10 nm to 30 nm and the remanent polarization changes from 0.25 C/m$^2$ for $h$=10 nm to 0.11 C/m$^2$ for $h$=30 nm [compare loops 1 – 5 in **Fig.4(a)**]. The polarization loop becomes rather slim and tilted with $h$ increase from 10 nm to 30 nm [compare loops 1 – 5 in **Fig.4(a)**]. The dielectric susceptibility loop shape changes from the one with two pronounced and relatively high maxima, which are well-resolved for $h$=10 nm, to the loop with rather low and smeared maxima, which are poorly resolved with $h$ increase up to 30 nm [compare loops 1 – 5 in **Fig.4(b)**]. Further increase of the film thickness $h$ does not lead to any changes in the loop shape and sizes, because for $h$ higher than 20 nm the loop itself is the dynamic effect, and the frequency decrease leads to its disappearance.

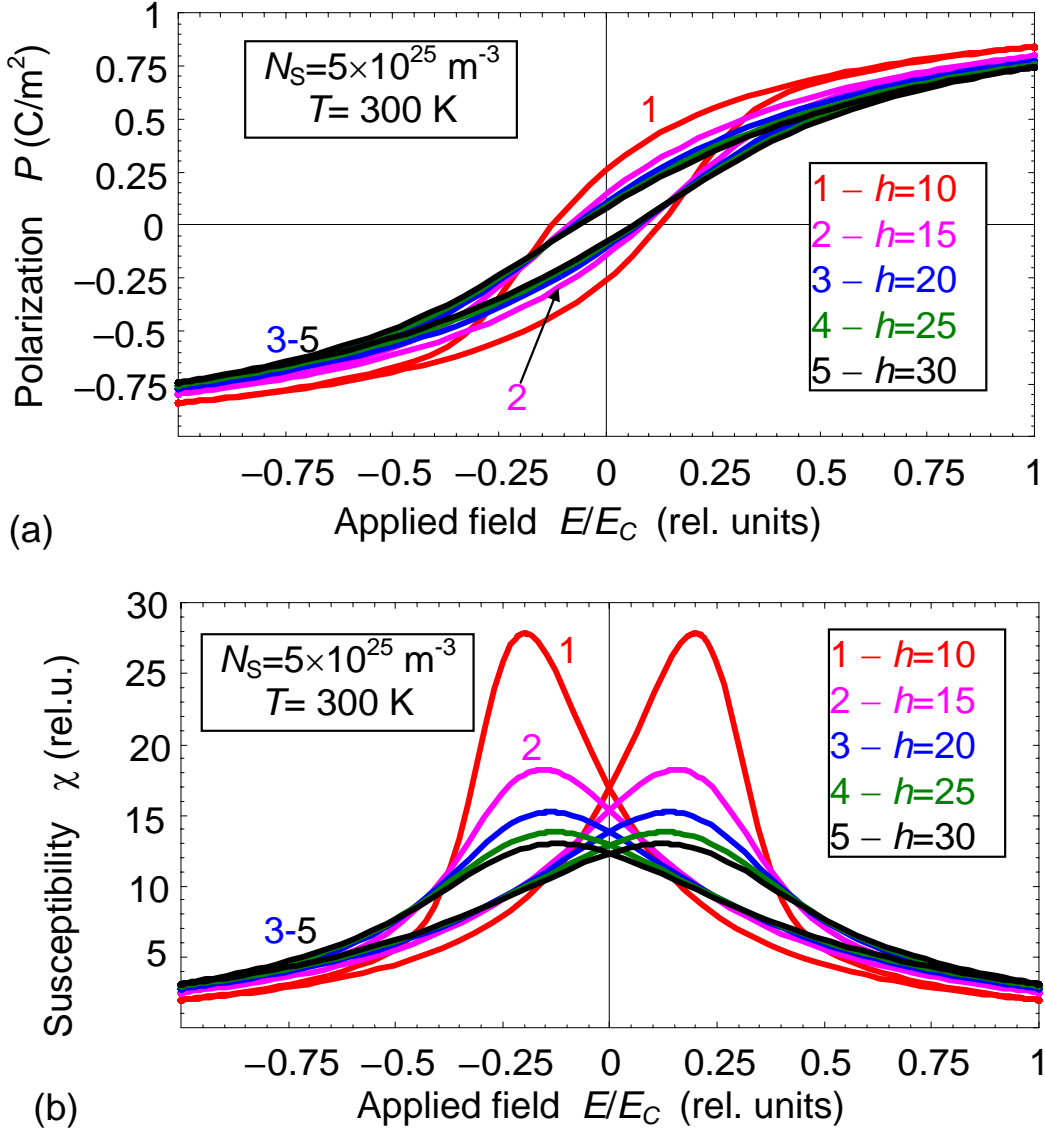

(a)

(b)

**FIGURE 4.** Polarization $P_s(T,h,N_s)$ (a) and dielectric susceptibility $\chi(T,h,N_s)$ (b) dependences on applied voltage calculated for room temperature T=300 K, $N_S = 5\times10^{25}$m$^{-3}$ and different film thickness $h$=(10, 15, 20, 25, 30) nm (curves 1, 2, 3, 4, 5). Dimensionless frequency $w$=0.05. Other parameters are the same as in **Fig.2**.

Notably, that polarization and dielectric susceptibility hysteresis loops shape, remanent polarization, coercive field, height and resolution of the dielectric susceptibility maxima, as well as the changes of these characteristic taking place with the film thickness increase are in qualitative agreement with experimental loops measured by Nishimura et al [16] and Polakowski and Muller [17]. The quantitative comparison of the proposed model with experiments [16, 17] is presented in the next section.

### B. Quantitative comparison with experiments

It follows from Eq.(11) that at fixed, e.g. room temperature $T = T_r$, the remanent polarization (that is proportional to the spontaneous one) can be represented as:

$$P_r(T = T_r, h) \cong P_{amp}\sqrt{\frac{h_{cr}}{h} - 1}, \qquad (13a)$$

where the polarization amplitude $P_{amp} = \sqrt{\frac{N_d}{P_d^2}\frac{k_B T_r}{\beta}}$. The critical thickness $h_{cr}$ in Eq.(13a) is given by expression

$$h_{cr}(T = T_r) = \left(-\frac{N_d p_d^2 \Lambda}{\varepsilon_0 \varepsilon_{el} k_B h_d T_r} - \frac{4 p_d^2 q_{33} W_{33} N_d N_S}{k_B T_r}\right) h_d. \qquad (13b)$$

Note that the critical thickness $h_{cr}$ linearly increases with the product of vacancies concentration and dipoles concentration $N_S N_d$, since the first term $-\frac{N_d p_d^2 \Lambda}{\varepsilon_0 \varepsilon_{el} k_B h_d T_r}$ is negative and the second one $-\frac{4 p_d^2 q_{33} W_{33} N_d N_S}{k_B T_r}$ is positive at $q_{33} W_{33} < 0$. The annealing in $N_2$ should increase the product $N_S N_d$, and so the expression (13b) explains the experimental result of Nishimura et al [16], who reported that the annealing increases of the thickness of ferroelectric $HfO_2$ film up to 35 nm.

The thickness dependence of the remanent polarization $P_r(T_r, h)$ is presented in **Fig.5(a)**. Diamonds are experimental points extracted from the *P-E* hysteresis loops measured by Polakowski and Muller [17], the red solid curve is calculated from Eqs.(13) with the fitting parameters $P_{amp} = 9$ μC/cm$^2$ and $h_{cr} = 12.3$ nm. Note the quantitative agreement of the solid curve calculated from Eqs.(13a) with experimental points in **Fig.5(a)**.

Hysteresis loops measured by Polakowski and Muller [17] also allow extracting the maximal polarization $P_{max}$ that is *h*-dependent [see boxes in **Fig.5(a)**]. Since the maximal polarization $P_{max}$ is also dependent on the amplitude $E$ of applied electric field, an interpolation for $P_{max}$ was found in the form:

$$P_{max}(T_r, h, E) \cong P_r(T_r, h, E) + \varepsilon_0 \chi(T_r, h) E \approx P_{amp}\sqrt{\frac{h_{cr}}{h} - 1} + \varepsilon_0 \chi(T_r, h) E, \quad (14)$$

The thickness dependences of the maximal polarization $P_{max}(T_r, h)$, calculated using Eq.(14) with the fitting function $\chi = \chi_b + \chi_m \sqrt{\frac{h_{cr}}{h} - 1}$ and parameters $\varepsilon_0 \chi_b E = 5.5 \, \mu\text{C/cm}^2$ and $\varepsilon_0 \chi_m E = 6 \, \mu\text{C/cm}^2$. Note the quantitative agreement between solid and dashed curves calculated from Eqs.(14) with experimental points in **Fig.5(a)**. However, the function $\chi = \chi_b + \chi_m \sqrt{\frac{h_{cr}}{h} - 1}$ in Eq.(14) is an interpolation that indicates the realization of the poly-domain scenario of polarization reversal in HfO$_2$ thin films.

For a single-domain scenario of the ferroelectric polarization reversal in thin films [59, 60] the "intrinsic" thermodynamic coercive field at fixed temperature $T_r$ is given by expression:

$$E_C(T_r, h) = \langle\langle \alpha_{33}^R \rangle\rangle \sqrt{\frac{27\langle\langle \alpha_{33}^R \rangle\rangle}{4\beta}} \cong E_0 \left(\frac{h_{cr}}{h} - 1\right)^{3/2}, \quad (15a)$$

where $E_0 = \sqrt{\frac{27}{4\beta}} \left(\frac{N_d}{P_d^2} \frac{k_B T_r}{\beta}\right)^{3/2}$ for the case of HfO$_2$ films with oxygen vacancies and zero frequency of external field, $\omega = 0$, since $E_C(T_r, h)$ is the static field by definition. The static intrinsic dielectric susceptibility is

$$\chi_S(T_r, h) = \frac{1}{\langle\langle \alpha_{33}^R \rangle\rangle + 3\beta P_S^2 + 5\gamma P_S^4} \sim \frac{1}{|(h_{cr}/h) - 1|}. \quad (15b)$$

Dotted curves in **Figs 5(b)** and **5(c)** are calculated from Eq.(15a) and Eq.(15b), respectively, and correspond to the "intrinsic" single-domain scenario of polarization reversal. They do not describe (even qualitatively) the experimental points (shown by diamonds) for the coercive field and susceptibility measured Polakowski and Muller [17].

For a poly-domain scenario the ferroelectric polarization reversal is conditioned by a domain nucleation and ruled by the motion of domain walls in the presence of multiple pinning centres (including the vacancies and/or other defects). Corresponding expression for the domain nucleation field $E_{DN}$ (that can be compared with the coercive field for real experiments) can be obtained as interpolation for the results of numerical modeling of the polarization reversal dynamics. For a considered case it appears proportional to the spontaneous (or remanent) polarization, $E_{DN}(T_r,h) \sim P_r \sim \sqrt{\frac{h_{cr}}{h}-1}$, and so obeys the expression:

$$E_{DN}(T_r,h) \cong E_{NS}\sqrt{\frac{h_{cr}}{h}-1}, \tag{16a}$$

For the poly-domain switching the expression for the relative dielectric permittivity $\varepsilon(T_r,h)$ corresponding to the zero-field point $E=0$ at the C-E hysteresis loop has the form

$$\varepsilon(T_r,h) = \varepsilon_b + \varepsilon_m\sqrt{\frac{h_{cr}}{h}-1}. \tag{16b}$$

Appeared that the expressions (16a) for the coercive field $E_{DN}(T_r,h)$ and (16b) for the relative dielectric permittivity $\varepsilon(T_r,h)$, shown by solid curves in **Fig.5(b)** and **5(c)**, respectively, describe quantitatively the experimental points [17] (shown by diamonds in **Fig.5(b)** and **5(c)**) and with the fitting parameters $E_{NS} = 1.3$ MV/cm, $\varepsilon_b = 23$ and $\varepsilon_m = 15$.

The discrepancies between the solid and dotted curves in **Fig.5(b)** and **5(c)** clearly represent the principal difference between the "intrinsic" single-domain and poly-domain scenarios of polarization reversal. Thus we meet with the poly-domain scenario of polarization reversal conditioned by a domain nucleation and ruled by the motion of domain walls in the presence of multiple pinning centres (including the vacancies and/or other defects).

Results presented in the **subsection III. B** allow us to conclude, that the developed theory describes semi-quantitatively the experimental results [16, 17].

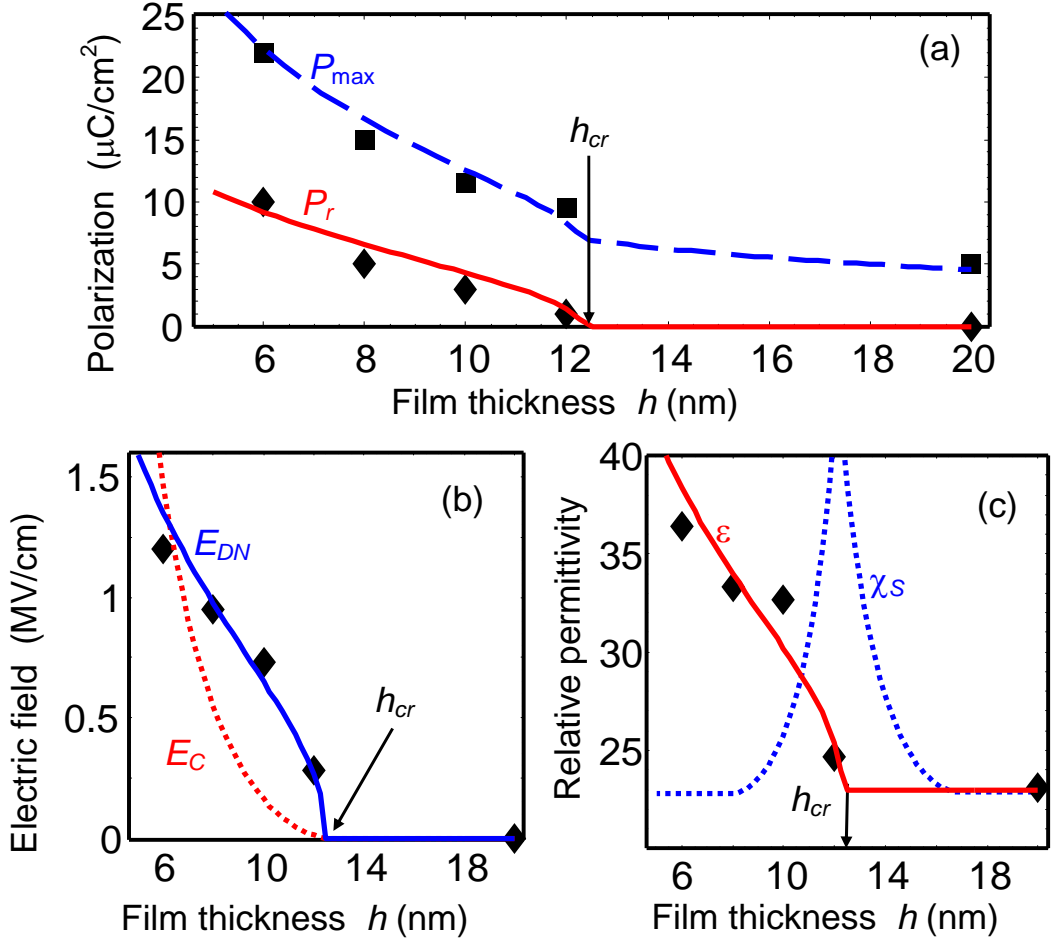

**FIGURE 5.** Thickness dependences of the **(a)** remanent ($P_r$) and maximal ($P_{max}$) polarization; **(b)** thermodynamic coercive ($E_C$) and domain nucleation ($E_{DN}$) fields; **(c)** relative dielectric permittivity ($\varepsilon$) and static susceptibility ($\chi_S$). Symbols are experimental points from Polakowski and Muller [17]. Solid curve for $P_r$ in **(a)** is calculated from Eq.(13a). Dashed curve for $P_{max}$ in **(a)** is calculated from Eq.(14). Dotted curve for $E_C$ in **(b)** is calculated from Eq.(15a). Solid curve for $E_{DN}$ in **(b)** is calculated from Eq.(16a). Dotted curve for $\chi_S$ in **(c)** is calculated from Eq.(15b). Solid curve for $\varepsilon$ in **(c)** is calculated from Eq.(16b). Fitting parameters are $P_{amp} = 9$ μC/cm², $h_{cr} = 12.3$ nm, $\varepsilon_0 \chi_b E = 5.5$ μC/cm² and $\varepsilon_0 \chi_m E = 6$ μC/cm² for plot **(a)**, $E_{NS} = 1.3$ MV/cm for plot **(b)**, $\varepsilon_b = 23$ and $\varepsilon_m = 15$ for plot **(c)**.

## IV. DISCUSSION

We proposed a model for the emergence of ferroelectric behavior in thin films of HfO$_2$ and related materials due to the electrochemical effects. Here, vacancies segregate at the interfaces described by surface concentration and decay length. The Vegard effect leads to the coupling between elastic dipoles and stresses, and further enables collective interactions that can give rise to ferroelectricity. Accumulation of electric dipoles in the vicinity of any inhomogeneities in the film, including both surfaces, lead to the appearance of long-range ordered ferroelectric polarization, while the elastic dipoles are the source of elastic long-range order in the structural phases, if there is a percolation. All the calculations were performed for the mixture of randomly distributed electric and elastic dipoles originated from oxygen vacancies being inherent defects in any oxide film (see **section II.A**).

The proposed model opens the way for calculation of binary oxides multiferroic properties and their complex phase diagrams. In accordance with experimental results [7, 16, 17] the diagrams of $HfO_2$ thin film include orthorhombic FE, cubic, tetragonal and monoclinic ferroelastic phases at room temperature. To perform the calculations of binary oxides thin film phase diagram one has to consider the polarization and elastic strain induced by oxygen vacancies as long-range order parameters of FE and ferroelastic phases, respectively. Keeping in mind the existence of magnetization in binary oxides thin film it can be considered as the ferromagnetic order parameter. In this case magnetoelectric and magnetoelastic interaction will represent the coupling between magnetization and two other long-range order parameters, and the coupling between them being realized by the Vegard-type interaction. It is not excluded that proposed approach will lead to conclusion about existence of multiferroicity not only in $HfO_2$ thin film, but also in the thin films of other binary oxides. In such a case one will face the unexplored and large enough group of room temperature multiferroic thin films. We argue that to verify the theoretically predicted decisive role of oxygen vacancies experimentally in multiferroicity origin, binary oxide films should be annealed of in oxygen atmosphere, by the same way as it was done earlier to check the influence of oxygen vacancies on magnetization [29-31].

Next let us discuss the possibility of electret-like polarization in thin films of binary oxides, because it may be useful for some applications. The Vegard-type elastic stress can originate from electrostriction (electrochemical coupling) or from surface-induced piezoelectricity (piezo-chemical coupling). As a result, ferroelectricity or electret-like state, respectively, can appear. In the case of surface-induced piezo-chemical coupling the electret-like irreversible polarization could exist. However the reversible ferroelectric polarization exists in $HfO_2$ thin films in accordance with experiment, speaking in favor of the electrochemical coupling domination.

## V. CONCLUSION

We have shown that oxygen vacancies segregation at interfaces, surfaces and grain boundaries can play the decisive role in the origin of ferroelectric properties of ultra-thin $HfO_2$ films and its polarization hysteresis. Specifically, the "electrochemical" coupling, that is the joint action of the omnipresent electrostriction and Vegard stresses originated from oxygen vacancies as elastic dipoles in the vicinity of surface.

All the calculations were performed in the assumption that oxygen vacancies, as elastic dipoles, can be partially transformed into electric dipoles due to the e.g. surface-induced piezoelectric effect. The electric dipoles are the sources of polarization coupled with elastic dipoles via the Vegard-type chemical stresses and electrostriction. Such electrochemical coupling leads to the sign change of the positive coefficient $\alpha$ in the term $\alpha P^2$ in the order-disorder type thermodynamic functional. Negative coefficient $\alpha$ becomes the driving force of the transition to the long-range ordered phase with the spontaneous polarization $P$ in the direction normal to the film surface. The proposed model allows describing the temperature and thickness dependences of spontaneous polarization, dielectric permittivity and hysteresis loops as a function of oxygen vacancies concentration.

In particular, we estimated that the reversible ferroelectric polarization, as high as $(0.05 - 0.2)$ $C/m^2$, can be induced by oxygen vacancies in $HfO_2$ films of thickness less than $(20 - 30)$ nm. Contrary the electret-like state, induced by the surface-induced piezoelectric coupling at distances less than 4 nm from each surface (arising due to the inversion symmetry breaking in the direction normal to the surfaces), the possible mechanism of ferroelectricity with reversable polarization appeared to be the electrochemical coupling existing entire $HfO_2$ thin films. The comparison of the theoretical results with experimental data obtained in the papers [16, 17] had shown, that the developed theory describes semi-quantitatively the experimental results. Hence, the proposed model of electric and elastic dipoles coexistence originated from the oxygen vacancies opens the way for quantitative description of multiferroicity and phase diagrams calculations in thin films of binary oxides.

Allowing for the generality of proposed approach it can be applied to any binary oxide, both undoped and doped ones. Furthermore, the proposed approach will be applicable to other cases of dipolar impurities, e.g. due to the trapped charges at the interfaces, etc.

**Authors' contribution.** M.D.G. and A.N.M. contributed equally to the research idea, problem statement and manuscript text. A.N.M. performed analytical and numerical calculations, and comparison with experiment, M.D.G., Y.K. and S.V.K. worked densely on the results interpretation and manuscript improvement.

**Acknowledgments.** Authors are grateful to Eugene A. Eliseev for useful remarks and stimulating discussions. A.N.M. work was partially supported by the Program of Fundamental Research of the Department of Physics and Astronomy of the National Academy of Sciences of Ukraine (project No. 0117U000240) and the European Union's Horizon 2020 research and innovation program under the Marie Skłodowska-Curie (grant agreement No 778070). S.V.K. research was supported by the Basic Energy Sciences, US Department of Energy.